\documentclass{webofc}

\usepackage[varg]{txfonts}   
\usepackage{hyperref}
\usepackage{url}
\usepackage[T1]{fontenc}
\hypersetup{colorlinks=true,citecolor=blue,urlcolor=blue,linkcolor=blue}
\begin{document}
\title{New developments in the Whizard event generator}
%
%

\author{\firstname{J\"urgen} \lastname{Reuter}\inst{1}\fnsep\thanks{\email{juergen.reuter@desy.de}} \and
        \firstname{Pia}
        \lastname{Bredt}\inst{2}\fnsep\thanks{\email{pia.bredt@uni-siegen.de}}
        \and
        \firstname{Marius}
        \lastname{H\"ofer}\inst{3}\fnsep\thanks{\email{marius.hoefer@kit.edu}}
        \and
        \firstname{Wolfgang}
        \lastname{Kilian}\inst{2}\fnsep\thanks{\email{kilian@physik.uni-siegen.de}}
        \and
        \firstname{Nils}
        \lastname{Kreher}\inst{2}\fnsep\thanks{\email{nils.kreher@uni-siegen.de}}
        \and
        \firstname{Maximilian}
        \lastname{L\"oschner}\inst{1}\fnsep\thanks{\email{maximilian.loeschner@desy.de}}
        \and
        \firstname{Krzysztof}
        \lastname{M\k{e}ka{\l}a}\inst{1,4}\fnsep\thanks{\email{krzysztof.mekala@desy.de}}
        \and
        \firstname{Thorsten}
        \lastname{Ohl}\inst{5}\fnsep\thanks{\email{ohl@physik.uni-wuerzburg.de}}
        \and
        \firstname{Tobias}
        \lastname{Striegl}\inst{2}\fnsep\thanks{\email{tobias.strieglr@uni-siegen.de}}
        \and
        \firstname{Aleksander Filip}
        \lastname{\.Zarnecki}\inst{4}\fnsep\thanks{\email{zarnecki@fuw.edu.pl}}
}

\institute{Deutsches Elektronen-Synchrotron DESY, Notkestr 85, 22607
  Hamburg, Germany
  \and
  Department of Physics, University of Siegen, Walter-Flex-Stra{\ss}e 3, 57068 Siegen, Germany
  \and
  Karlsruhe Institute of Technology, Institute for Theoretical
  Physics, Wolfgang-Gaede-Str. 1, 76131 Karlsruhe, Germany
  \and
  Faculty of Physics, University of Warsaw, Pasteura 5, Warszawa,
  02-093, Poland
  \and
  University of W\"urzburg, Institut f\"ur Theoretische Physik und
  Astrophysik, Emil-Hilb-Weg 22, 97074 W\"urzburg, Germany
}

\abstract{We give a status report on new developments within the
  Whizard event generator. Important new features comprise NLO
  electroweak automation (incl. extension to BSM processes like
  SMEFT), loop-induced processes and new developments in the UFO
  interface. We highlight work in progress and further plans, such as
  the implementation of electroweak PDFs, photon radiation, the
  exclusive top threshold and features for exotic new physics
  searches.}
\maketitle
\section{Introduction}
\label{intro}

Whizard is a multi-purpose Monte Carlo event generator which was
released in its first version 25 years ago, with the main focus on
simulating physics at lepton colliders, especially linear
colliders. It has been greatly modernized due to the needs of LHC
physics~\cite{Kilian:2007gr} towards version 2 in 2010, and then
underwent another drastic development for the next-to-leading order
(NLO) automation for version 3 in
2021~\cite{Binoth:2009rv,Greiner:2011mp,Weiss:2017qbj,ChokoufeNejad:2017rag,Rothe:2021sml,Stienemeier:2022wmy,Bredt:2022nkq,Bredt:2022dmm}.
Whizard comes with its own 
tree-level matrix element generator,
O'Mega~\cite{Moretti:2001zz,ChokoufeNejad:2014skp,Ohl:2023bvv}.
Whizard supports fully general beam polarization in terms of spin
density matrices, factorized processes in narrow-width approximation
with full spin correlations, comes with an analytic parton
shower~\cite{Kilian:2011ka}, and directly interfaces to shower and
hadronization of Pythia6~\cite{Sjostrand:2006za} or
Pythia8~\cite{Sjostrand:2014zea} (other shower and hadronization
programs are possible through LHE files). Whizard uses an adaptive
multi-channel Monte Carlo integration (VAMP)~\cite{Ohl:1998jn} with 
phase-space parameterizations optimized for either resonant or
multi-peripheral multi-particle processes which also has an
MPI-parallelized incarnation~\cite{Brass:2018xbv}.

Strong emphases in Whizard is laid upon the simulation of
lepton-collider beam spectra via the dedicated subpackage
CIRCE2~\cite{Ohl:1996fi}. This uses a 2-dimensional binned histogram
fit to the beam spectrum with a special smoothening and a dedicated
treatment of the high-energy peak to avoid artifical beam energy
spreads.

Whizard supports a special exclusive treatment of the top threshold,
matching the NRQCD next-to-leading logarithmic (NLL) corrections with
the relativistic NLO QCD
corrections~\cite{ChokoufeNejad:2016qux,Bach:2017ggt}.

For BSM models, Whizard supports the most recent UFO2
standard~\cite{Darme:2023jdn}, while still keeping partial backwards
compatibility via its Feynrules
interface~\cite{Christensen:2010wz}. Some recent applications of the 
UFO2 interface are shown
here~\cite{Han:2021lnp,Mekala:2022cmm,Mekala:2023diu,Mekala:2023kzo,Celada:2023oji}.

In the next section, we are summarizing new features and ongoing
developments roughly since the last LCWS in May 2023 at SLAC.


\section{New developments and current progress}

As there are several proposed collider options for electron-positron
colliders, Whizard has provided several CIRCE2 spectra for using their
beam spectra in simulations. Since quite some time, ILC and CLIC
spectra have become available while CEPC was provided around the time
of the CEPC CDR. Recently, beam spectra for the Cool Copper Collider
(C${}^3$) have been simulated; these simulations have been shown that
a special handling for depleted regions within the spectra is needed
in order to avoid the appearance of unphysical artifacts. Simulations
for the FCC-ee spectra are coming soon, however, the beam delivery
system of FCC-ee is still being updated after the feasibility study
midterm report. Also, due to the work of the SLAC group around Timothy
Barklow on XFEL-based lasers, photon-collider options have been
revived, which can also be simulated with CIRCE2. A special
Gaussian-like spectrum will become available in the future as
well. End of 2023, support for muon collider spectra have been added
to CIRCE2, but as of now no particle  muon collider spectrum is
available.

For QCD quantum numbers, Whizard uses the color flow basis for their
representation and evaluation~\cite{Kilian:2012pz}. Recently, the
matrix element generator O'Mega has been geeneralized for arbitrary
tensors in color space (epsilon tensors, sextets, decuplets, higher
exotic representation, which are potentially important for colorful
dark-sector models)~\cite{Ohl:2024fpq}. The support of these exotic
color structures in the Whizard interface is work in progress, to be
released in v3.2.

One active line of development is the completion of NLO EW corrections
at lepton colliders with massless fermions, while those for massive
fermions are already completed~\cite{Bredt:2022nkq}. The current
implementation of NLL QED PDFs has been completed in
2023~\cite{Reuter:2023vei}, while in the past months the numerical
stability has been greatly improved close to the integrable
singularity $z\to 1$. Complete NLO EW results will be ready within the
next few months, which will then allow to also properly match (soft-
and collinear) photon radiation to all orders, where first hard-coded
attempts have been made in~\cite{Kilian:2006cj,Robens:2008sa} without
solving the problem of double-couting. For the very high-energy
regime, electroweak factorization might become relevant, e.g. for muon
colliders in the multi-TeV regime or FCC-hh. This approach is similar
to the equivalent vector-boson approximation (EVA or EWA) which,
however, is a purely kinematic approximation, where the corresponding
factorized piece is not necessarily a splitting function obeying a
DGLAP equation. Those objects that fulfill a DGLAP equation are rather
called EW PDFs. While the traditional implementation of the EWA in
Whizard has been recently rederived and revalidated, the basic
infrastructure for EW PDFs has been implemented as well. This will
allow to read in EW PDFs from LHAPDF-like grids, to be interpolated
and DGLAP-evolved to according scales. Here, we do expect first
reliable numerical results mid-2025.

Another current line of development is to port Whizard to GPUs;
already at LCWS last year it was reported that there is an
automated way to port O'Mega matrix elements as CUDA code to be
evaluated on the GPU. Meanwhile, the VAMP Monte-Carlo integrator
routines have been ported to the GPU as well with a medium number of
sanity checks and a as of yet small number of benchmarking runs. These
look quite promising but are not yet conclusive as for simple
phase-space configurations the efficiency is dominated by the matrix
elements. Work in progress is to also offload cut expressions and
clustering statements to the GPU, and first non-trivial are to be
expected early next year. 

\section{Summary, open developments and outlook}

This is a status report on the availability of new features in the
Whizard event generator based upon the version 3.1.5 from late summer
2024. Besides a large number of bug fixes on the UFO interface, on the
NLO interfaces to the one-loop providers, the event format interfaces
to HepMC and for intermediate diagram restrictions in matrix
elemenets, there was support for Pythia versions 8.310 and
newer. Among the new features is the extension of the resonance-aware
FKS NLO subtraction both for hadron-collider processes as well as to
NLO EW corrections. Also the GoSam interface has been overhauled to
support NLO processes at LHC. The current development is preparing
interfacing the upcoming new major release 3 of GoSam and full support
for NLO QCD corrections for UFO-based BSM models. Finally, the special
simulation for photon-induced processes to low-$p_T$ hadrons at low
energies, parameterized with cross sections measured at fixed-target
experiments like Crystal Ball that had been used for the ILC TDR
simulations with Whizard 1.95 has now been ported to Whizard 3.1.5.
The current main lines of development include the finalization of
(squared) matrix elements exclusive in coupling orders in conjunction
with a refactoring of quantum number handling in Whizard, the GPU
parallelization of the full Whizard integration (and ultimately also
event generation) and also fully general color structures. The NLO
development focuses on NLL QED PDFs for complete NLO EW cross sections
at lepton colliders with massless fermions, in the further future a
soft (YFS) exponentiation of photons, and the support for effective W
approximation or NLO EW PDFs. Also, a native EDM4HEP interface is
being developed. A next update is expected for the FCC Physics
Workshop in January 2025.

\section*{Acknowledgments}

For this work we would thank the organizers for a fantastic conference
with stimulating atmosphere. For different parts of Whizard
development, we acknowledge funding by the Deutsche
Forschungsgemeinschaft (DFG, German Research Foundation) under grant
396021762 - TRR 257, under grant 491245950 and under Germany's
Excellence Strategy-EXC 2121 ``Quantum Universe"-390833306, by
National Science Centre (NCN, Poland) under the OPUS research project
no. 2021/43/B/ST2/01778 and by EAJADE - Europe-America-Japan
Accelerator Development and Exchange Programme (101086276), and by
the International Center for Elementary Particle Physics (ICEPP), the 
University of Tokyo.

\bibliography{whizard_jrr}

\end{document}